\documentclass{llncs}

\usepackage{fullpage}
\usepackage{times}
\usepackage{graphicx}
\usepackage{subfigure}
\usepackage{moreverb}
\usepackage{epsfig}
\usepackage[normalem]{ulem}

\def\elegans{{\it{C. elegans}~}}

\begin{document}

\title{Characterizing Self-Developing Biological Neural Networks: A First Step Towards their
Application To Computing Systems}
\date{\today}

\author{
Hugues Berry, Olivier Temam\\
\{hugues.berry,olivier.temam\}@inria.fr\\
}
\institute{Alchemy-INRIA Futurs\\Parc Club Orsay Université\'e, ZAC des vignes,
4 rue Jacques Monod,
91893 Orsay Cedex France}

\maketitle



\begin{abstract}
Carbon nanotubes are often seen as the only alternative technology
to silicon transistors.  While they are the most likely short-term
alternative, other longer-term alternatives should be studied as
well, even if their properties are less familiar to chip designers.
While contemplating biological neurons as an alternative component
may seem preposterous at first sight, significant recent progress in
CMOS-neuron interface suggests this direction may not be
unrealistic; moreover, biological neurons are known to self-assemble
into very large networks capable of complex information processing
tasks, something that has yet to be achieved with other emerging
technologies.\\

The first step to designing computing systems on top of biological
neurons is to build an abstract model of self-assembled biological
neural networks, much like computer architects manipulate abstract
models of transistors and circuits.  In this article, we propose a
first model of the {\it structure} of biological neural networks. We
provide empirical evidence that this model matches the biological
neural networks found in living organisms, and exhibits the {\it
small-world} graph structure properties commonly found in many large
and self-organized systems, including biological neural networks.
More importantly, we extract the simple local rules and
characteristics governing the growth of such networks, enabling the
development of potentially large but realistic biological neural
networks, as would be needed for complex information
processing/computing tasks.  Based on this model, future work will
be targeted to understanding the evolution and learning properties
of such networks, and how they can be used to build computing
systems.
\end{abstract}

\section{Introduction \label{introduction}}

Carbon nanotubes look like a promising alternative technology to
silicon chips because the manufacturing process, possibly based
upon self-assembly, will be much cheaper than current CMOS
processes. On the other hand, these individual components may turn
out to be much slower than current transistors, exhibit lots of
manufacturing defects, and may be difficult to assemble into
complex and irregular structures like today's custom processors.
Current research are focused on building increasingly large
structures of carbon nanotubes and understanding how they can be
transformed into computing devices.

However, carbon nanotubes, though the most promising and short-term,
is not the only possible alternative to silicon chips. Other emerging
technologies, even if they are less familiar to chip designers, should
be explored as well. In this article, we focus our attention on {\it
biological neurons}. They share some properties with carbon nanotubes:
they have a low design cost, but they will provide even slower
components, a significant percentage of these components will be
similarly faulty, and it will be hard to assemble them into complex,
irregular {\it pre-determined} structures. On the other hand, they have a
significant asset over carbon nanotubes: we already know it is
possible to self-assemble them into very large structures capable of
complex information processing tasks.

While proposing computing structures based on biological neurons may
seem preposterous at first sight, G. Zeck and P.
Fromherz~\cite{Fromherz2003,Zeck2001} at the Max Planck Institute
for Biochemistry in Martinsried, Germany, have recently demonstrated
they can interface standard silicon chips with biological neurons,
pass electrical signals back and forth through one or several
biological neurons, much like we intend to do with carbon nanotubes,
i.e., hybrid carbon nanotures/standard CMOS
chips~\cite{goldstein-isca01}. Moreover, based on this research
work, Infineon (one of the main European chip manufacturers) has
recently announced it is investigating a prototype of a chip (called
"NeuroChip") that can interconnect a grid of transistors with a
network of biological neurons~\cite{WebInfineon}, based on
Fromherz's research work. So, while we will not claim this research
direction should be mainstream, it is certainly worth exploring.

Now, computing machines, such as current processor architectures,
are designed using a very abstract model of the physical
properties of transistors and circuits. Typically, what processor
architects really use (e.g., at Intel or other chip manufacturers)
is how many logic gates can be traversed in a single clock cycle,
and how many logic gates can be laid out on a single chip. They do
not deal with the complex physics occurring at the transistor
level, they rely upon a very abstract and simplified model of the
undergoing physical phenomena. Similarly, if we want to start
thinking about computing systems built upon biological neurons, we
must come up with sufficiently abstract models of biological
networks of neurons that will enable the design of large systems
without dealing with the individual behavior of biological
neurons.

The vast literature on artificial neural networks provides little
indications on the {\it structures} of biological neural
networks~\cite{Haykin1999}. To understand what kind of computing
systems can be built upon biological neurons, we must first
understand the kind of {\it structures} into which biological
neurons can self-assemble. Consequently, we have turned to biology
for that issue, and the current article is a joint work between
computer science and biology research groups.  We start with the
biological neural network of a small living organism, a worm named
{\it Caenorhabditis elegans}, which has been described in great
details in~\cite{Albertson1976,White1986}.\footnote{This worm has
been subject to intense scientific study as one of the most simple
living organism that retains many of the characteristics of complex
organisms, such as a brain, learning capabilities, and other
physiological similarities. As a side note, it is the first organism
which genome has been entirely sequenced.}  Based on this work,
Oshio {\it et al.}~\cite{Oshio2003} have recently built a database
which describes this biological neural network and facilitates its
manipulation. Using this map as an oracle, we define a model of
network growth in real space and provide empirical evidence that the
characteristics of networks built upon this model and the above
mentioned biological network closely match. Since this model
describes the network {\it growth} using simple local rules, it can
be used to represent much larger networks, as would be needed for
computing systems. Note that such models had not yet been derived by
and are not directly useful to biologists: there are many studies on
biological neural networks, but they focus on the identification of
{\it regular} biological networks with clear structures, such as the
basic circuit of the visual cortex~\cite{Douglas1998}, and they do
not account for the seemingly irregular structure of the vast
majority of biological neural networks. We provide a {\it
statistical} description of these apparently unstructured biological
networks, that can be used as a building block for computing systems
studies. Future work will focus on analyzing the evolution and
learning properties of neural networks with such structures.

In Section~\ref{part1}, we present the biological neural network of
\elegans and study its properties. In Section~\ref{part2}, we
build a network model with similar properties, provide empirical
evidence that it closely emulates the neural network of \elegans, and
provide a detailed comparison of the model and its biological
counterpart.

\section{A biological neural network \label{part1}}

\begin{figure}
  \center
  \includegraphics[scale=0.45]{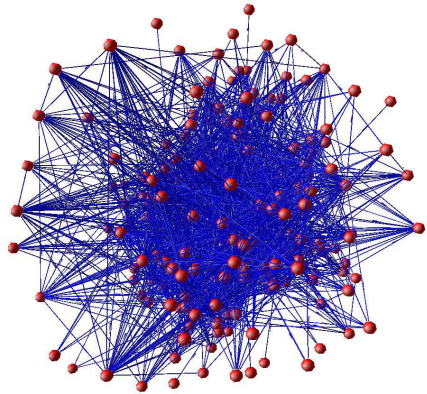}\\
  \caption{Visual illustration of the neural network of \elegans. Neurons are displayed as nodes and
  connections between them are symbolized as links. Spatial positions are arbitrary.}\label{figure1}
\end{figure}

\elegans is a small (millimetric) worm with a simple network of
302 neurons. All the connections between its neurons have been
mapped~\cite{Albertson1976,White1986} and are believed to be
relatively well conserved between individual worms. To construct a
graph model of this system, we used the electronic database
recently published by Oshio {\it et al}.~\cite{Oshio2003}. A part
of this system, comprising 20 neurons and referred to as the
"pharyngeal system" is dedicated to control rhythmic contractions
of a muscular pump that sucks food into the worm
body~\cite{Albertson1976}. This system is almost totally
disconnected from the rest of the network. Following Morita {\it
et al.}~\cite{Morita2001}, we neglected here the pharyngeal system
and only deal with the remaining 282 neurons. We then further
neglected those neurons for which no connection had been
described, as well as the connections to non specified cells. At
the end, the network thus consisted of 265 neurons. Unlike Morita
{\it et al.}~\cite{Morita2001}, we treated each link as directed,
i.e., we differentiated links from neuron $i$ to neuron $j$ and
links from $j$ to $i$; however, we collapsed multiple identical
links into a single one. In opposition to chemical synapses, which
are unidirectional connections, some of the neuron connections,
called electrical synapses or {\it gap junctions}, are
bidirectional. Here we treated gap junctions as pairs of links
with opposite directions. Overall, we obtain 2335 unique links (or
10234 connections if we allow redundant links with the same
orientation between two neurons).

\begin{figure}
  \center
  \includegraphics[scale=0.85]{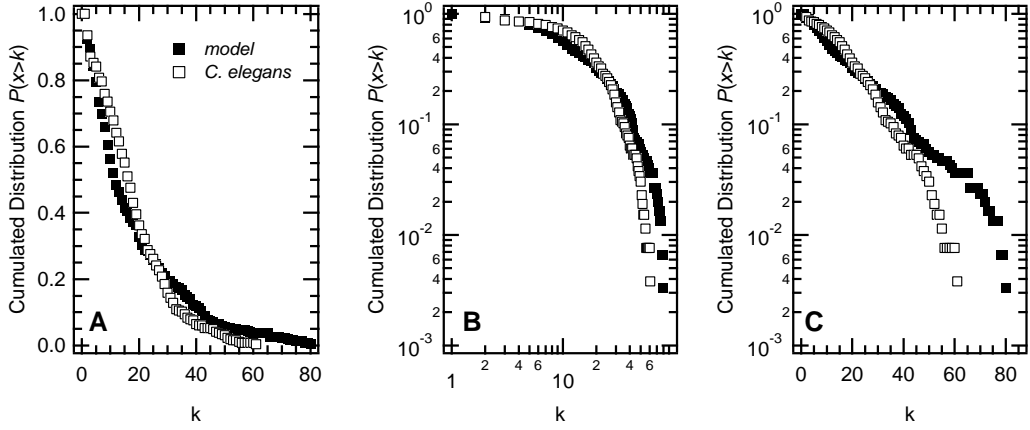}\\
  \caption{Cumulated distributions of the connectivity, $P(x>k)$, where $k$ is the connectivity (i.e., the number of links to and from each node), for the \elegans neural network (white squares); the black squares curve corresponds to the network model later introduced and commented in Section~\ref{part2}. The same data are presented as
  a linear-linear (A), a log-log (B) and log-linear plot (C).}\label{figure2}
\end{figure}

Figure~\ref{figure1} shows a visual illustration of the
corresponding neural network. A visual inspection of this figure,
especially the peripheral nodes,\footnote{On paper, the core of the
network structure is barely visible, but on a screen, it can be
inspected through zooming and 3D manipulations; however the
peripheral structure is the same as the core structure.} indicates
that the network is rather heterogeneous: strongly connected nodes
coexist with sparsely connected ones.  We further tried to estimate
the nature of the probability distribution of the connectivity (or
graph degree), as it plays a fundamental role in characterizing the
network type.  The probability distribution of the connectivity in
\elegans neural network has been controversial. A first study
claimed the distribution was compatible with a power-law (graphs
determined by power-law distributions are also called ``scale-free''
graphs)~\cite{Barabasi1999}. Not long after, this result was
contradicted by an article from H.E. Stanley's team that studied
outgoing and incoming connectivity separately (and ignored gap
junctions) and showed that both distributions were exponential, thus
excluding scale-free properties~\cite{Amaral2000}. Finally, Morita
{\it et al.} put forward correlations among incoming, outgoing and
gap junctions to explain that the total degree (incoming + outgoing
+ gap junctions) was neither exponential nor displayed a clear power
law decrease~\cite{Morita2001}. Figure~\ref{figure2} presents the
distribution of the total connectivity for \elegans neural network
(white squares).  The center panel is a replot of the left one, in
log-log coordinates. A power law decrease would yield a straight
line in this representation, which is clearly not the case. Further,
the right panel is another replot of the same data, in log-linear
coordinates. Here, a straight line would indicate an exponential
decrease. A clear exponential decrease is not apparent from this
panel. Thus, our results confirm that connectivity distribution for
\elegans neural network is neither scale-free nor clearly
exponential.

We will see in the next section that additional network
characteristics are necessary to emulate this network structure; more
importantly, we will extract the simple local rules governing the
network growth, enabling the development of potentially large but
realistic biological networks using the same rules.

\section{A model of biological neural networks \label{part2}}

{\bf Small-World graphs and neural networks.} The global behavior
of most large systems emerges from local interactions between
their numerous components. At an abstract level, these systems can
often be viewed as graphs, with each link representing the
interaction between two components. Such graph theory approaches
have proven successful in understanding the global properties of
several complex systems originating from highly disparate fields,
from the biological to social and technological domain. Hence the
same (or similar) reasonings can be applied to understand cell
metabolism~\cite{Jeong2000}, the citation of scientific
articles~\cite{Newman2001}, software
architecture~\cite{Valverde2002}, the Internet~\cite{Barabasi1999}
or electronic circuits~\cite{Ferrer2001}. A common feature of all
these networks is that their physical structure reflects their
assembly and evolution, so that their global features can be
understood on the basis of a small set of simple local rules that
control their growth. The most common statistical structures
resulting from these local rules are the so-called
\textit{small-world} and \textit{scale-free} networks. Small-world
properties characterize networks with both small average shortest
path and a large degree of clustering, while scale-free networks
are defined by a connectivity probability distribution that
decreases as a power law (more formal definitions will be given in
the following). At a much coarser grain, graph theory methods have
recently been applied to networks of cortical
areas~\cite{Sporns2004,Eguiluz2005},\footnote{Cortical areas are
functionally related zones of the cortex that contain
approximately $10^8$ neurons} i.e., not networks of neurons but
networks of neuron {\it areas}, with the prospect of understanding
the network functions.  Since we target the characterization of
networks of biological {\it neurons}, we study the neural network
of the millimetric worm \elegans at the level of individual {\it
neurons}, and attempt to derive a network growth model that
closely emulates it.

Most complex networks can be categorized into four
families~\cite{Barabasi2004}. In {\it random networks} (also known
as Erd\H{o}s-R\'{e}nyi graphs), two nodes $i$ and $j$ are
connected with a predefined probability independently of all
others. These graphs are characterized by short paths between two
nodes (denoted $\lambda$) and a low clustering (denoted $\langle C
\rangle$). On the opposite, {\it regular} graphs (where each node
has the same connectivity) are characterized by a high clustering
and a large average shortest path. Between these two extremes,
{\it small-world} graphs display both small average shortest paths
and a high degree of clustering. For most small-world networks,
$P(k)$, i.e., the probability distribution of the connectivity
$k$, decreases very quickly (exponentially) beyond the most
probable value of $k$, which thus sets the connectivity scale.
However, in some graphs (such as the Internet), $P(k)$ decreases
as a power-law of $k$ ($P(k) \propto k^{-\gamma}$), i.e., in a
much slower way. In this case, nodes with a very high connectivity
(hubs) can also be present with a significant probability so that
the connectivity does not display a clear scale, hence the term
``scale-free'' networks.

We now formally introduce the parameters of a network model. Besides
the number of nodes $N$ and number of links $K$, the structural characteristics
of complex networks are mainly quantified by their link density
$\rho$, average connectivity $\langle k \rangle$, connectivity
distribution $P(k)$, average shortest path $\lambda$ and average
clustering coefficient $\langle C \rangle$~\cite{Sporns2002}. The
network density $\rho$ is the density of links out of the $N(N-1)$
possible directed links\footnote{Each node can have at most $N-1$ outgoing
links, so the maximum number of links is $N(N-1)$.} (recall multiple links
between two nodes are considered a unique link and self-connections
are forbidden)
\begin{equation}\label{eq1}
   \rho=K/\left(N^2-N\right)
\end{equation}
The connectivity (or \textit{degree}) $k_i$ of node number $i$ is
the number of links coming from or directed to node $i$. $P(k)$ is
the probability distribution of the $k_i$'s and $\langle k
\rangle$ their average over all the nodes in the network. Let
$d(i,j)$ be the shortest path (in number of neurons) between
neuron $i$ and $j$, then $\lambda$ is its average over the network
\begin{equation}\label{eq2}
    \lambda=1/\left(N^2-N\right)\sum_{i,j}d(i,j)
\end{equation}
The clustering coefficient of a node $i$ with $k_i$ (incoming plus
outgoing) connected neighbors is defined by
\begin{equation}\label{eq3}
    C_i=E_i/\left(k_i^2-k_i\right)
\end{equation}
where $E_i$ is the number of connections among the $k_i$ neighbors
of node $i$, excluding the connections between a neighbor and node
$i$ itself. The average clustering coefficient $\langle C \rangle$
is the average of the $C_i$'s over all nodes and expresses the
probability that two nodes connected to a third one are also
connected together (degree of cliquishness).

The main structural characteristics of the \elegans neural network
are indicated in Table~\ref{Table1}. Compared to a random network
with the same density, this neural network has a similar average
shortest path but the clustering has increased almost fivefold. This
means that, in the \elegans neural network, one neuron can reach any
other neuron in only three connections on average. This is a clear
sign of small-world properties. Considering the network is treated
here as a directed graph, these results are coherent with previously
published estimates~\cite{Watts1998,Clark2003}.

\begin{table}
  \begin{center}
  \begin{minipage}{5cm}
  \begin{tabular}{|c|c|c|c|c|}
    \hline
    Network & $\rho$ &$\langle k \rangle$ & $\lambda$&$\langle C \rangle$\\
    \hline
    \elegans & 0.033 & 17.62 & 3.19 & 0.173 \\
    {\it random} & 0.033 &  17.62 & 2.75 & 0.0352 \\
    {\it model} & 0.030 & 17.73 & 3.37&0.175 \\ \hline\hline
  \end{tabular}
  \newline
  \end{minipage}
  \end{center}
  \caption{Structural characteristics of the neural network of \elegans shown in Figure~\ref{figure1},
  a comparable random (Erd\H{o}s-R\'{e}nyi) network and
  the network obtained with
  the proposed growth model. The data for the random network are averages over 20 network realizations.
  See text for definition of
  the listed properties.}\label{Table1}
\end{table}

{\bf In biological neural networks, distance matters.}  In order to
design a network growth model of the \elegans neural network that
achieves small-world properties, we have found that taking into
account the physical {\it distance} between two nodes (neurons) is
critical.  Most network growth models do not consider this
parameter~\cite{Kimura2004}. For instance, most scale-free networks
are obtained through a preferential attachment rule which postulates
that new nodes are linked to the already most connected
nodes~\cite{Barabasi1999}. Not only this development rule implies
some global control mechanism (i.e., a node must somehow know which
are the most connected nodes) which seems unlikely in the case of a
neuronal system, but it also implies that long connections are just
as likely as shorter ones. Similar arguments can be opposed to the
Watts-Strogatz rewiring algorithm that generates small-world
networks through addition of long-range connections to a
pre-existing regular circular network~\cite{Watts1998}. An improved
variation of the Watts-Strogatz algorithm restricts rewiring to a
local spatial neighborhood around each node~\cite{Clark2003} thus
implicitly introducing the distance factor. However, these last two
models are not and cannot evolve into network growth models and thus
do not provide a biologic realistic metaphor. In opposition to these
models, we address in the present work the specific case of
biological neuron network \textit{growth} in real three dimensional
space.

Several observations in biology further support the key notion of
physical distance. Long distance connections are expensive in
biological neural networks because they imply large volumes of
metabolically active tissue to be maintained and long transmission
delays~\cite{Chklovskii2002}, just like links between two nodes in
Internet or in airport transportation systems are more costly.
Moreover, wiring length optimization seems to be a crucial factor of
cortical circuit development~\cite{Cherniak1994,Cherniak1995}. The
network structure itself depends on the wiring length. For instance,
small-world properties (as well as, under some circumstances,
scale-free connectivity~\cite{Ferrer2003}) have been shown to emerge
naturally upon minimization of the euclidian distance between
nodes~\cite{Mathias2001}. Furthermore,
Kaiser~et~al.~\cite{Kaiser2004} have recently shown that network
structure during growth in a metric space is influenced by neuron
density (number of neurons per unit volume) when growth occurs in a
spatially constrained domain.

{\bf A network growth model for \elegans.}  We now propose a
network growth model in a three-dimensional space. Neurons are
abstracted as cubical volumes of unit size. The position of each
neuron on the cubic lattice is defined by the integer coordinates
$(i,j,k)$ of its center of mass and spans over the volume
comprised between $(i-1/2, j-1/2, k-1/2)$ and $(i+1/2, j+1/2,
k+1/2)$. The lattice dimensions are $L_x$, $L_y$ and $L_z$,
defining a volume of $L_x \times L_y \times L_z$ unit sizes. Each
step of the growth algorithm consists of six elementary substeps:
\begin{enumerate}

\item choose a neuron $n$ at random among the neurons already
connected in the network (\textit{origin} neuron). Let $(i,j,k)$
be the spatial coordinates of $n$ on the lattice. \item Then
choose a \textit{destination} site $(i',j',k')$ at distance $d$
with probability
\begin{equation}\label{eq4}
    P(d)= 1/\xi\exp(-d/\xi)
\end{equation}
where $d$ is the euclidian distance between $(i,j,k)$ and
$(i',j',k')$, and $\xi$ is a parameter that sets the average
connection distance.  If the chosen destination site is located
outside the lattice borders, go back to substep 1. Thanks to the
exponential distance distribution, the probability to create a
connection of a given length (at a certain distance) decreases
rapidly with the length, which accounts for wiring minimization.
Note that, in biological neural networks, new connections are
established through cell outgrowths (neurites) from existing
neurons; these outgrowths are guided by gradients of chemical
concentration which similarly decay rapidly with distance. We also
tested other probability distributions, such as power-laws, and
obtained similar results.

\item If a neuron $n'$ of the network already exists at the
destination site
    $(i',j',k')$, a connection is created between $n$ and $n'$.

\item If there is no neuron at the destination site, a new neuron
$n'$ is placed at the destination site
    (and a connection is created between $n$ and $n'$) with
    probability $P_{new}$; the value of $P_{new}$ is discussed
    below.

\item If a connection has been created during one of the two
preceding steps, its direction ($n \rightarrow n'$ or $n'
\rightarrow n$) is chosen with probability $P_{n \rightarrow
n'}=k_n^{out}/\left(k_n^{out}+k_n^{in}\right)$ where $k_n^{out}$ and
$k_n^{in}$ are respectively the outdegree and indegree of neuron
$n$, i.e., the number of connections with $n$ as starting node and
$n$ as destination node. This probability distribution reflects the
property that, in \elegans, strongly connected neurons (hubs)
exhibit asymmetric fractions of incoming/outgoing connections:
either much more incoming or much more outgoing connections. The
{\it joint degree distribution matrix} of the \elegans network shown
in Figure~\ref{figure4}A highlights this characteristic: each matrix
element $m_{ij}$ corresponds to the number (coded using a gray
scale) of neurons with $i$ incoming and $j$ outgoing connections, so
that a scattered plot indicates an asymmetrical repartition of
incoming and outgoing connections for many neurons; hubs are the
dots located farthest from the origin (lots of connections). We can
observe that several of the hubs are located far from the diagonal,
hence the asymmetry.

\item go back to substep 1.

\end{enumerate}

This algorithm iterates until the network contains a prescribed
number of neurons $N$. In this study, we set $N=300$.

\begin{figure}
  \center
  \includegraphics[scale=0.95]{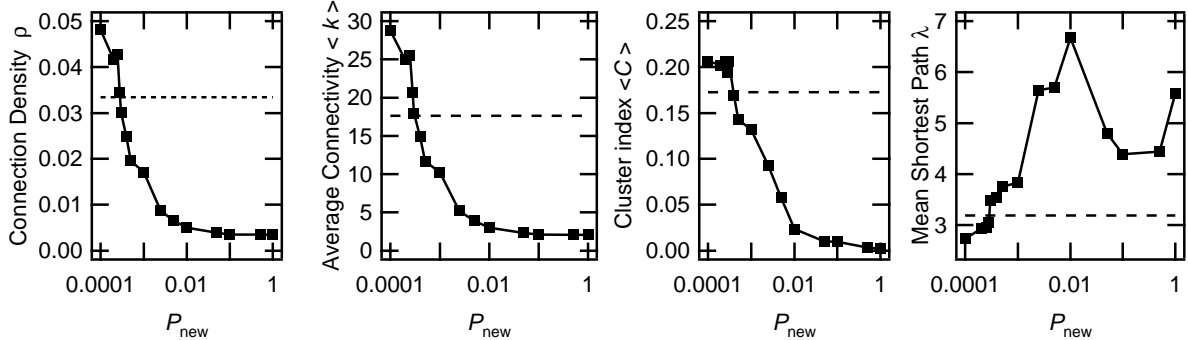}\\
  \caption{Structural characteristics of the networks obtained
  with the proposed model as a function of the probability
  $P_{new}$ that a new neuron connects to the network. The
  dashed line indicates the corresponding value found for the
  network of \elegans. Other parameters were $L_x \times L_y
  \times L_z = 51
  \times 51 \times 51$
  (neuron size units), $\xi = 10$, $N=300$}\label{figure3}
\end{figure}

\begin{figure}
\center
  \includegraphics[scale=0.65]{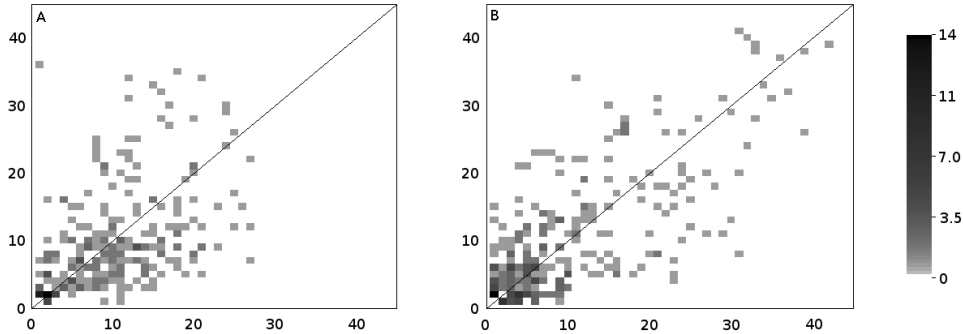}\\
  \caption{Joint degree distribution matrices for the neural network of \elegans (A) and a model neural network obtained with $P_{new}=0.00028$ (B).
  Each element $\{m_{ij}\}$ of the matrices is color-coded with a
grey level proportional to the number of neurons displaying $i$
incoming and $j$ outgoing links, as shown in the colorbar on the
right. The straight line simply aims at highlighting the symmetry
axis denoted by the diagonal. All network parameters are the same
as in Figure~\ref{figure3}.}\label{figure4}
\end{figure}

{\bf New neurons are unlikely to be created in already cluttered
areas.} If one sets $P_{new}=1$, i.e., a new neuron is certain to
be created in a currently empty location, our algorithm is
essentially a three-dimensional extension of the model recently
proposed by Kaiser and Hilgetag~\cite{Kaiser2004}. As demonstrated
by these authors, networks obtained with $P_{new}=1$ progressively
acquire small-world properties when neuron density approaches 1
(i.e. when $N\rightarrow L_x \times L_y \times L_z$). In this case
however, the average connectivity and the connection density (as
well as the joint degree distribution matrix, see below) remains
severely smaller than observed in \elegans (e.g., we could not
achieve an average connectivity higher than 6-7).

Decreasing $P_{new}$ breeds significantly more realistic results. As
seen in Figure~\ref{figure3}, connection density, average
connectivity and clustering index increase as $P_{new}$ decreases
while the average shortest path $\lambda$ decreases. Thus,
decreasing $P_{new}$ yields networks with increasingly strong
small-world properties together with increasingly high average
connectivity. Furthermore, Figure~\ref{figure4}B indicates that our
model indeed yields hubs with asymmetric connectivity, as observed
for \elegans in Figure~\ref{figure4}B. Taken together, the results
of Figure~\ref{figure3} and~\ref{figure4} show that all the studied
structural properties of the networks obtained with our algorithm
match that of \elegans neural network for $P_{new} \approx 3 \times
10^{-4}$. A biological interpretation of this value is that natural
neural networks would be very reluctant to admit new neurons in the
network (as only 1 contact out of $\approx$3000 would be
statistically successful). Interestingly, recent results in
neurobiology suggest that the lack of neural turnover and/or
replacement of injured neurons in the adult brain is not due to the
absence of potentially competent cell, but, more probably, to a
strong reluctance of the neurons to accept newcomers into an already
established neural network~\cite{Rakic2004}. In light of these
findings, our results suggest that this strong reluctance could be
one of the factors inducing the high average connectivity observed
in biological neural networks, i.e., if new neurons can hardly
emerge in already cluttered areas, connections are mostly drawn
among existing neurons, hence the high connectivity. Since we made
the observation that $P_{new}$ must be very small while fine-tuning
the model and before being aware of these recent neurobiological
findings, we interpreted this correlation as additional evidence of
the validity of our model.

Finally, we can note that the model network characteristics match
the \elegans network characteristics, as shown in
Table~\ref{Table1}. And, in spite of some discrepancies due to the
relatively small number of neurons for high connectivity values, the
connectivity distribution of the model network is also fairly close
to that of the \elegans network, as shown in Figure~\ref{figure2},
see the black squares curve. In particular, the model network
distribution is neither a power-law, nor clearly exponential (though
it is closer to an exponential decrease than that of \elegans). We
can thus conclude that our model closely emulates the structure of
the biological network of \elegans.

\section{Conclusions \& Future Work}

Based on empirical data of a tiny organism, we have elaborated a
model for biological neural networks. In agreement with previous
works~\cite{Watts1998}, we have found that the neural network of
\elegans has a graph structure with {\it small-world} properties,
like many complex systems found in nature. Because the model
defines the network {\it growth} properties, we can now use it to
characterize the large neural networks needed for achieving
computing tasks. The next step will consist in studying how this
network structure affects the learning capabilities and
characteristics of neural networks. We then intend to pursue two
research directions.

A further step will consist in improving the model accuracy/realism by
integrating known but abstract characteristics of the behavior of
individual neurons. Finally, through this combined model, we will
investigate the application of such biological neural networks to
computing tasks, assuming the experimental setups described
in~\cite{Fromherz2003}. In this perspective, our aim is to obtain a sufficiently abstract model of biological
networks of neurons that will enable the design of large systems
without dealing with the individual behavior of biological
neurons. We think that the availability of such abstract models will be a crucial chokepoint that will have to be overcome if we want to build
computing systems using real biological neurons.

In parallel, we intend to use the model developed in this article for
biology-oriented studies, especially to investigate how evolution
combined with such network structures can foster the emergence of new
simple functions within a neural network.

\bibliographystyle{plain}

\small

\bibliography{shared}

\end{document}